\documentclass[aps,prl,reprint,groupedaddress,showpacs,amsmath,amssymb,twocolumn,floatfix]{revtex4-1}
\usepackage{graphicx}
\usepackage{bm}
\usepackage{color}
\usepackage[normalem]{ulem}
\usepackage{cleveref}
\usepackage{amsmath}

\newcommand{\CF}{$_{1}^{2}$CF }
\newcommand{\CFs}{$_{1}^{2}$CFs }

\begin{document}

\title{Evidence for pairing states of composite fermions in double-layer graphene}

\author{J.I.A. Li$^{1}$}
\email{J.I.A.Li and Q. Shi contributed equally to this work}
\author{Q. Shi$^{1}$$^{*}$}
\author{Y. Zeng$^{1}$}
\author{K. Watanabe$^{2}$}
\author{T. Taniguchi$^{2}$}
\author{J. Hone$^{3}$}
\author{C.R. Dean$^{1}$}

\affiliation{$^{1}$Department of Physics, Columbia University, New York, NY, USA}
\affiliation{$^{2}$National Institute for Materials Science, 1-1 Namiki, Tsukuba, Japan}
\affiliation{$^{3}$Department of Mechanical Engineering, Columbia University, New York, NY, USA}

\date{\today}



\maketitle

\textbf{Pairing interaction between fermionic particles leads to composite Bosons that condense at low temperature. Such condensate gives rise to long range order and phase coherence in superconductivity, superfluidity, and other exotic states of matter in the quantum limit. In graphene double-layers separated by an ultra-thin insulator, strong interlayer Coulomb interaction introduces electron-hole pairing across the two layers, resulting in a unique superfluid phase of interlayer excitons.  In this work, we report a series of emergent fractional quantum Hall ground states in a graphene double-layer structure, which is compared to an expanded composite fermion model with two-component correlation. The ground state hierarchy from bulk conductance measurement and Hall resistance plateau from Coulomb drag measurement provide strong experimental evidence for a sequence of effective integer quantum Hall effect states for the novel two-component composite fermions (CFs), where CFs fill integer number of effective LLs ($\Lambda$-level). Most remarkably, a sequence of incompressible states with interlayer correlation are observed at half-filled $\Lambda$-levels, which represents a new type of order involving pairing states of CFs that is unique to graphene double-layer structure and beyond the conventional CF model.}



Within the narrowly dispersing landau levels (LLs) that define the quantum Hall effect (QHE) regime, the kinetic energy is quenched. The resulting  electron behaviour is therefore determined almost entirely by minimizing Coulomb repulsion. This results in the series of correlated states appearing at fractional LL filling, known as the fractional quantum Hall effect (FQHE) ~\cite{Tsui_FQHE,Laughlin1983}. In double-layer quantum wells consisting of closely spaced parallel 2DEGs, even richer QHE physics emerges. In the small separation limit, the additional layer degree of freedom and interlayer Coulomb interactions lead to a variety of new correlated states that are tunable with interlayer separation and transverse displacement fields. Experimentally observed examples include formation of a superfluid exciton condensate ~\cite{Kel.04,Tutuc.04,Nandi.12,Eis.14,Li.17a,Liu.17a} between electrons in one layer and holes in the other, occurring at total integer LL filling (half filling in each layer) and even denominator FQHE states at 1/2 and 1/4 total filling ~\cite{Halperin1983,Suen1992,Eisenstein1992,Suen1994,Shabani2013}.  Theoretical work has identified a host of other possible states at fractional total filling, some of which are expected to be exotic non-Abelian states with topologically non-trivial excitations ~\cite{Wen1992, Scarola2001,Alicea2009,Ardonne2002,Wen2010, Zaletel.15,EunAhKim2015}.  However, compared to single layer systems the FQHE in double layers has been less explored experimentally, and many of these states remain unobserved.  

Here we report measurement of the FQHE in dual-gated double-layer graphene (DLG) heterostructures where the active regions consist of two graphene monolayers separated by a layer of hexagonal boron nitride (hBN) (see SI for details of the device structure). Several recent efforts have demonstrated that in DLG, the  hBN spacer can be made as thin as 1-2~nm before interlayer tunneling becomes relevant ~\cite{Britnell2012tunnel,Gor.13,Li.16,Tutuc.16,Li.17a,Liu.17a}. Consequently the effective interlayer separations $d/\ell_{B}$, which characterizes the interlayer coupling strength, can remain less than $0.5$ for $B$-field up to $30$ T.  This provides access to a previously unexplored regime in quantum Hall bilayers that combines strong intralayer (high $B$) and strong interlayer (small effective $d$) Coulomb repulsion, and in a device structure where the layer densities can be independently tuned. We identify insulating interlayer states with a Corbino geometry, which recently was shown to yield improved resolution of FQHE sates in single layer graphene devices ~\cite{Zeng2018}, and confirm the nature of these states with Coulomb drag measurements in a Hall bar geometry. At large magnetic fields, the Corbino geometry reveals a rich sequence of incompressible ground states that has no analogue in monolayer or Bernal-stacked bilayer graphene. We interpret this sequence in the context of the composite Fermion (CF) model for two-component systems ~\cite{Scarola2001, Jain.03}, finding excellent agreement with this construction.  Additionally we observe several states that fall outside of two-component CF construction but which can be understood as the result of pairing interaction between CFs.

\begin{figure*}
 \includegraphics[width=0.9\linewidth]{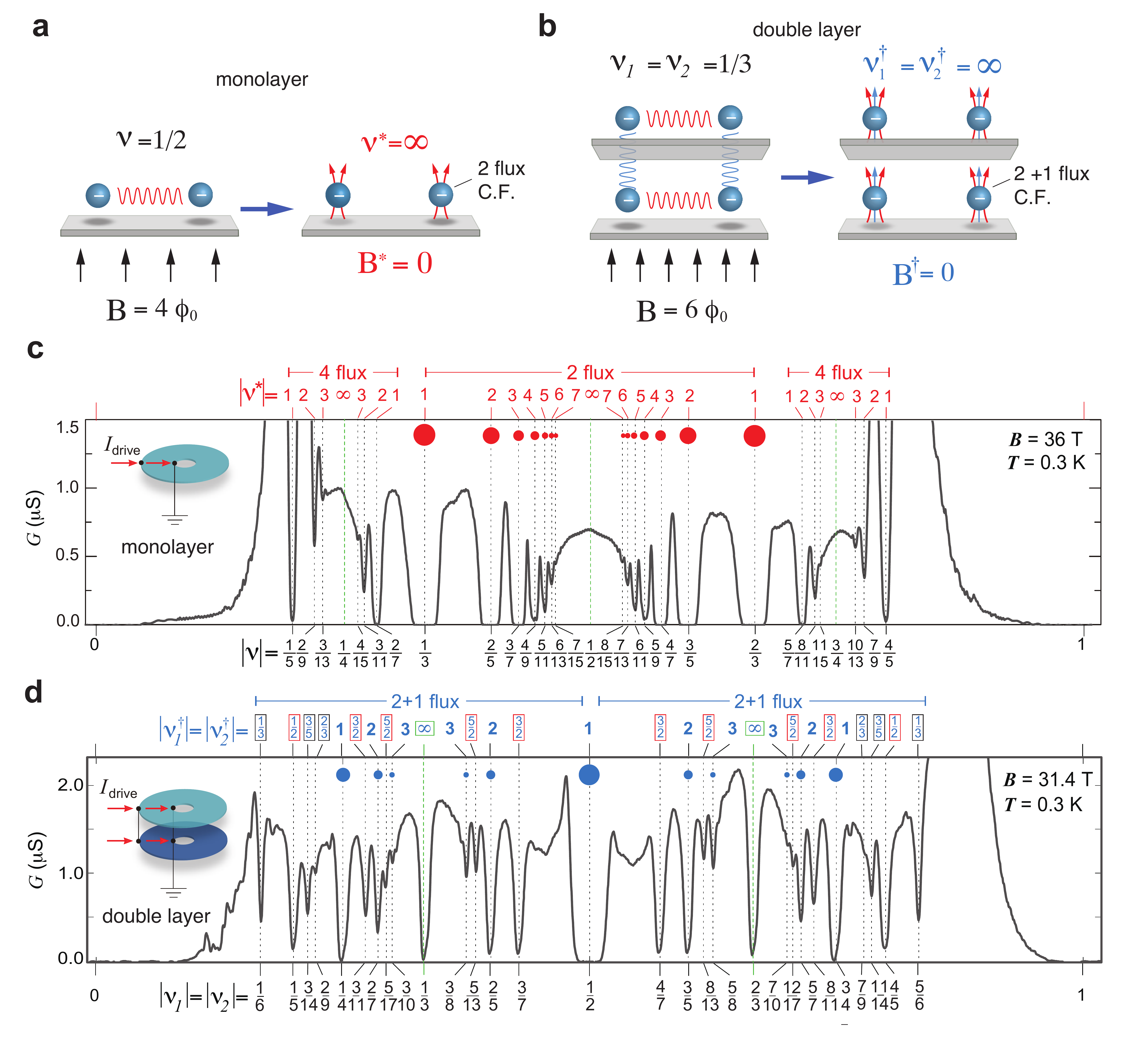}
 \caption{\label{fig1}{\bf{Composite fermion construction and bulk conductance measurement in Corbino geometry.}} 
 Electron interactions within a landau level can be modeled by attaching magnetic flux tubes to electrons to form non-intearction CFs. This process also modifies the effective magnetic field experienced by CFs, leading to an effective filling fraction that is different from the bare electrons. Zero effective field for CFs corresponds to LL filling $\nu=1/2$ for single layers (a) and $\nu_{1}=\nu_{2}=1/3$ for double layers (b) (see text). 
 (c) Bulk conductance, $G$, versus $\nu$, measured in a single graphene Corbino disk. The effective filling fraction, $\nu^{*} $, for the single layer 2-flux and 4-flux CFs are indicated in red on the top axis. (d) $G$ versus $\nu_{1}=\nu_{2}$ for a double layer Corbino, measured at equal layer density. The corresponding effective filling, $\nu^{\dagger}_{i}$,  for 2+1-flux CFs are indicated in blue along the top axis (see text). Red and blue circles in (c) and (d), respectively, highlight minima corresponding to integer valued effective CF filling in each system.  The circle radius is drawn proportional to the width of the conductance minimum. Inset in c and d show the Corbino measurement geometry.
 }
\end{figure*}

In the single component CF model, an even number of flux quanta are attached to each electron,  transforming the strongly interacting electrons into a system of nearly independent CF quasi-particles (this transformation is illustrated in Fig. 1a).  Owing to the flux attachment, each CF also moves in a reduced effective magnetic field, $B^{\ast}=B-an\phi_0$, where $n$ is the carrier density, $\phi_{o}$ is the magnetic flux quantum and $a$ is the number of flux quanta attached to each electron.  At filling fraction $\nu=1/a$ the effective magnetic field is precisely zero and the CFs behave as a metal with a well defined Fermi surface ~\cite{Halperin1993}. Away from these Fermi surfaces, the FQHE states at fractional electron filling are reinterpreted instead as effective IQHE states of CFs, where the effective CF filling fraction, $\nu^{\ast}$, is related to the real electron filling, $\nu$,  by the relation $\nu^{\ast}=\nu/(1-a\nu)$ where $a$ is the flux attachment number. This remarkably simple construction makes it possible to interpret a wide range of complex behaviours associated with the correlated FQHE states within the context of a non-interacting single particle picture ~\cite{Jain.03}.

Fig. 1c shows bulk conductance versus filling fraction, measured in a monolayer graphene Corbino disk at $B=36$~T.  The data range spans a single branch of the lowest LL between $\nu=0$ to $\nu=1$. A large number of FQHE states are visible ( fractional denominators as large as 15 are resolvable),   confirming the excellent sample quality and transport resolution achievable in the Corbino geometry. The top axis in Fig 1c labels the  filling fraction of the CF LLs  (referred to as $\Lambda$-levels) calculated from the above relation for the 2-flux ($a=2$) and 4-flux ($a=4$) series.  The  sequence of states and their hierarchy are in excellent agreement with the non-interacting CF model ~\cite{Scarola2001,Jain.03}.

For double layer systems,  an expanded two-component CF model has been proposed to account for both intra-- and interlayer types of interactions ~\cite{Scarola2001,Jain.03}. In addition to the flux attachment scheme described above, an additional number of flux quanta penetrating one layer are attached to electrons of the opposite layer (we illustrate this as blue arrows in Fig. 1b).  This construction renormalizes both the intra-- and interlayer interactions, once again resulting in a system of nearly non-interacting CF quasi-particles.  For this case the effective magnetic field seen by the CFs is determined by the electron density in both graphene layers, $B_{i}^{\dagger} = B - (a n_i+ b n_j)\phi_0$, where the subscript $i$ denotes layer index, $a$ and $b$ are the number of intralayer and interlayer flux attachment, respectively. We note several important features of this transformation: (i) The CFs retain their layer index, but the layers become nearly independent of one another; (ii) The CFs can experience different effective magnetic fields when the layer densities are not matched (iii) While the intralayer flux attachment must be an even number, the interlayer flux can be any integer value (but no larger than a). In this work however we will focus on CFs with $a=2$ intralayer and $b=1$ interlayer flux attachment, which we refer to as a $2+1$-flux CF or \CF for simplicity.

 
Fig. 1d shows similar measurement but for a double layer Corbino device with $d=2.7$~nm interlayer spacing (this measurement is obtained in the matched density condition by flowing current through the two layers simultaneously, as depicted in the cartoon inset). We plot the bulk conductance vs individual layer filling fraction over the same range as in Fig. 1c, so that in the absence of interlayer interactions the two plots should be identical.   However, it is clear upon inspection that the presence of the second layer substantially modifies the response. While again we resolve a large number of FQHE states (labelled by the individual layer filling fraction on the bottom axis), the hierarchy of states shows no correlation to the single layer response, and novel structures appear such as the condensate of electron-hole pairs at $\nu_1=\nu_2=\frac{1}{2}$ ~\cite{Li.17a,Liu.17a}.
 
\begin{figure*}
\includegraphics[width=0.9\linewidth]{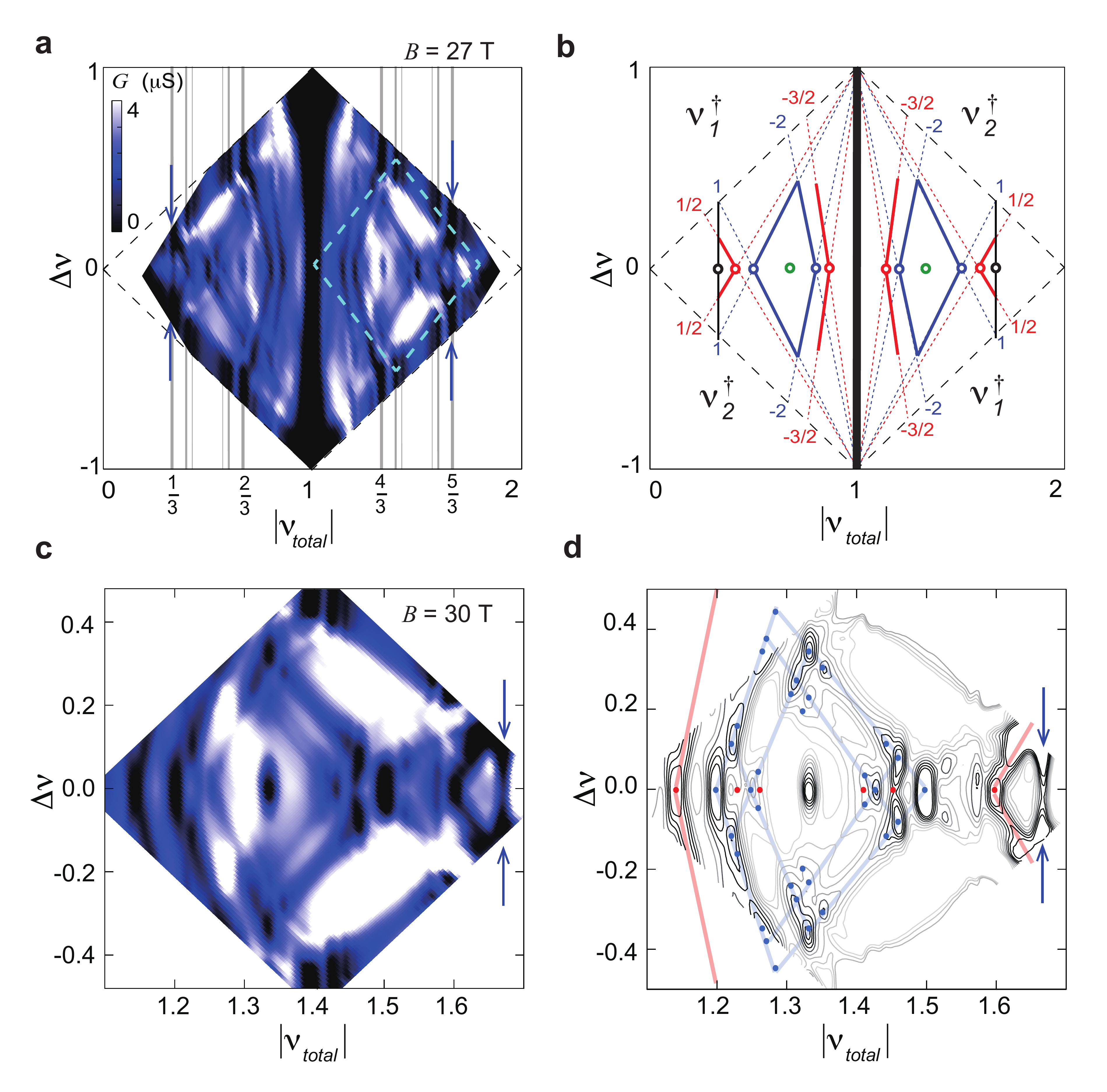}
\caption{\label{fig4} {\bf{Density imbalance}} 
(a) $G$ versus $\nu_{total}$ and $\Delta\nu$ measured at $B = 27$ T. (b) Schematic phase diagram showing constant integer and half integer CF filling in each graphene layer as dashed blue and red lines, respectively. Features observed in bulk conductance are highlighted as solid blue, red and black lines. The colored circles correspond to insulating features in Fig.~1d under matched density condition. Blue, red and black circles mark integer, even denominator and odd denominator $\nu^{\dagger}$ states, respectively, whereas green circles correspond to  $\nu_i = \frac{1}{3}$ and its particle-hole conjugate, where \CFs form a Fermi surface phase.   (c) High resolution measurement of the region highlighted by a dashed boundary in (a), acquired at $B = 30$ T. (d) Contour plot of bulk conductance data shown in (c).  Blue dots correspond to effective integer \CF filling  in both graphene layers. Red dots mark the effective half filling states.   Blue arrows point to the feature corresponding to the $(333)$ state.   }
\end{figure*}  
 
 
\begin{figure*}
\includegraphics[width=1.00\linewidth]{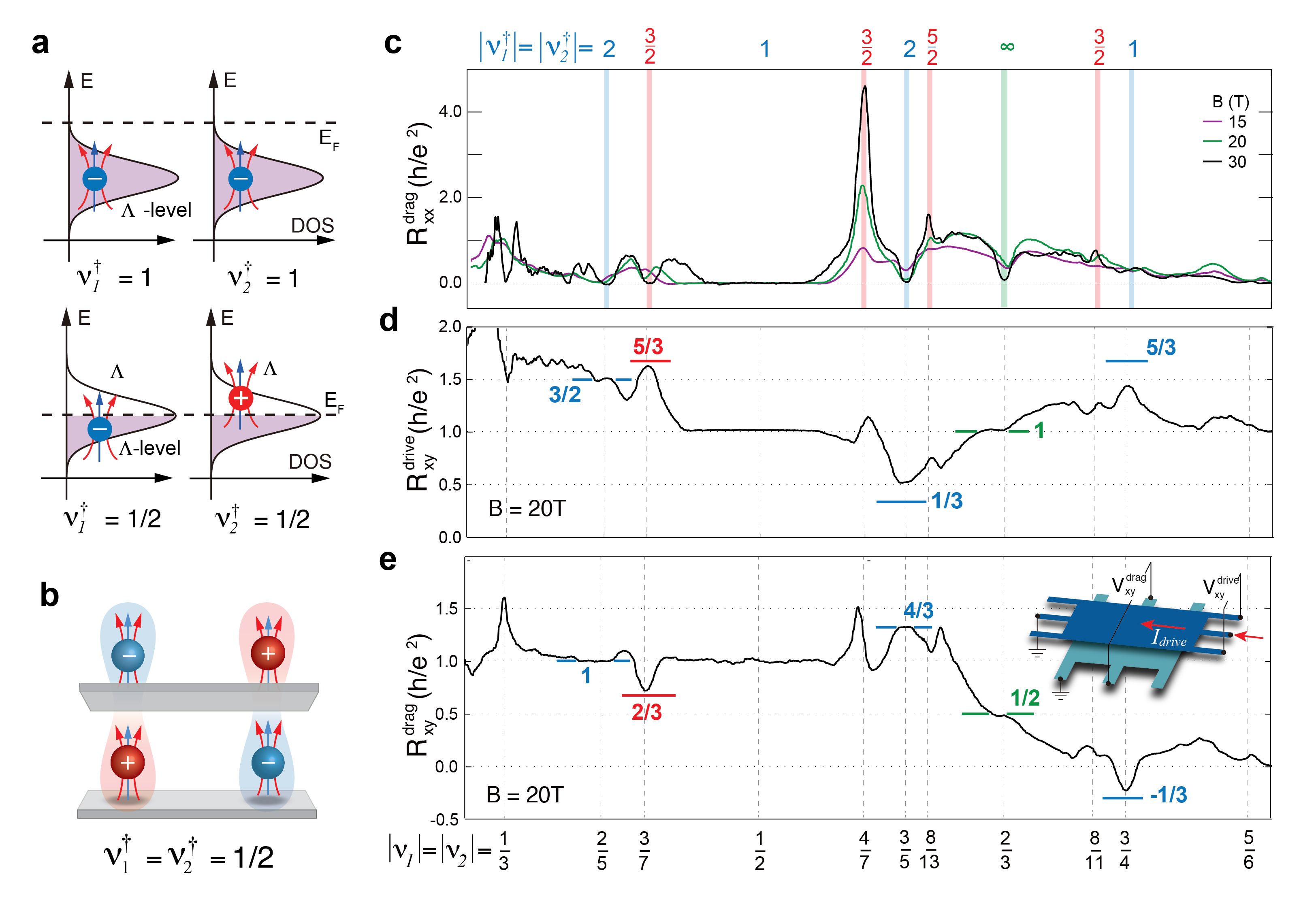}
\caption{\label{fig3} {\bf{Coulomb drag measurement in Hall bar geometry}} Schematics $\Lambda$-levels for different effective CF filling. (a) Integer CF filling for $CF_2^1$ in both layers. (b) Half integer CF filling for $CF_2^1$ in both layers. (c)-(d) $R_{xy}^{drive}$, $R_{xy}^{drag}$ and $R_{xx}^{drag}$ versus $\nu_i$ along equal density line measured at $B = 20$ T and $T = 0.3$ K. Coulomb drag is performed on a device with interlayer separation $d=2.5$ nm. Inset of (e), schematic of Coulomb drag geometry (see method for more discussion) . }
\end{figure*} 
 
In the two-component \CF model, the effective $\Lambda$-level filling fraction at matched layer density is given by~\cite{Scarola2001,Jain.03}
\begin{equation}
\nu^{\dagger}_{i}=\nu^{\dagger}_{j}=\frac{\nu}{1-3\nu},\label{1}\\
\end{equation}
\noindent
where $i$ and $j$ label the layer index, and we use superscript ''$^{\dagger}$'' to distinguish from the single component CF model.  In the top axis of Fig. 1d we label the \CF filling fractions for each of the observed FQHE states. A prominent sequence of insulating features corresponding to integer values of $\nu^{\dagger}_{i}$ is evident, and its hierarchy are in excellent agreement with the two-component \CF model, converging to the expected Fermi surface ($\nu^{\dagger}_{i}=\infty$) at electron LL filling $\nu_{i}=1/3$ and $2/3$ (blue circles in Fig. 1d). 
However, a larger number of additional FQHE states are also present that do not correspond to integer filling of \CF $\Lambda$-levels. These fall into three distinct categories: (i) fractional $\nu^{\dagger}$ with even denominator (highlighted by red squares in Fig. 1d) (ii) fractional $\nu^{\dagger}$ with odd denominator (black squares) and (iii) $\nu^{\dagger}=\infty$ ($\nu=1/3$ and $2/3$) where a Fermi surface of CFs is expected (green squares).  
We note that incompressible states at partially filled $\Lambda$-level, i.e. outside of CF IQHE sequence, presumably involve correlation between CFs. An example of this kind of behaviour has been reported in GaAs single layers where for example the $\nu=\frac{4}{11}$ state was argued to be a 1/3 FQHE of CFs, resulting from residual interactions between the CF quasi-particles ~\cite{Pan.03}. In double layers, residual CF interactions play a more prominent role, stabilizing a wider variety of ground states with strength comparable to integer $\nu^{\dagger}$ ~\cite{Scarola2001,Jain.03}. The relative strength of fractional and integer $\nu^{\dagger}$ states indicates an effective Coulomb interaction different from a single 2D confinement, due to the interlayer interaction. 

Our dual-gated geometry allows us to independently tune the density of each layer and therefore map the evolution of these states away from the layer-balanced condition. 
Fig.~2a plots bulk conductance as a function of total electron filling fraction,  $\nu_{tot} = \nu_{1}+\nu_{2}$, and difference between the individual layer filling fractions,  $\Delta\nu = \nu_{1}-\nu_{2}$.   In the chosen color scale dark blue indicates low conductance (either one or both layers is weakly conducting),  while white indicates high conductance. Fig. 2b summarizes the main features of Fig. 2a where the solid lines identify prominent trajectories of low conductance.  Open circles in Fig. 2b label the same  layer-balanced \CF states that were identified in Fig. 1d, where blue, red and black indicated integer, even denominator, and odd denominator  $\nu^{\dagger}_{i}$ states, respectively.   Fig.~2c  plots a higher resolution map of the dashed region in Fig. 2a, and Fig. 2d shows a corresponding contour plot where the sequence of incompressible states inside this dashed area can be seen in more clear detail.  

For density imbalance between the layers, equation 1 can be generalized to 
\begin{equation}
\nu_{i}^{\dagger}=\frac{\nu_{i}}{1-2\nu_{i}-\nu_{j}}.\label{2}\\
\end{equation}
\noindent
The dashed lines in Fig. 2b reflect lines of constant individual layer CF filling, according to equation 2.  
We observe that states appearing at integer $\nu^{\dagger}_{i}$ values under density balance evolve along  trajectories (solid blue lines) that  match the expected trajectories for integer filling of \CF states in one layer (dashed blue lines).
The intersection point between dashed blue lines, highlighted by blue circles in Fig.~2d, corresponds to regions where both layers are expected to have integer-valued $\nu^{\dagger}_{i}$ according to equation 2, and therefore both layers should be insulating. Fig.~2c and d demonstrate robust insulating features at each of these crossing points, and show a well defined hierarchy as judged by the depth of the conductance minimum: namely, the most robust ground states are observed at the four corners of the diamond shaped area, where $\Lambda$-level filling is the lowest and the effective magnetic field is the strongest. The observed trajectories of the integer $\nu^{\dagger}_{i}$ states with layer imbalance, and their hierarchical behavior, provide further evidence that these states are well described by the \CF model. 

Among the non-integer $\nu^{\dagger}_{i}$ states, the even denominator states also evolve along lines of constant filling (solid red lines in Fig. 2b), matching the expected trajectories for half-filled \CF $\Lambda$-levels (dashed red lines).  
We note that along the red line trajectories, one of the layer remains at constant half-integer $\Lambda$-level filling, while the other layer varies over a large range of effective filling.  A state persisting along this trajectory therefore could indicate pairing between CFs within the half-$\Lambda$ filled layer only, and with no correlation to the other layer. We speculate that these states may be of a similar origin to the Pfaffian that is believed to describe the half filling even denominator state in single layer systems ~\cite{Scarola2001,Jain2012,Jain2014}, though from the Corbino data alone we cannot definitively confirm this. 
At the matched density condition, two copies of the presumed Pfaffian could persist, i.e. one in each layer, although interlayer correlations between the \CFs could also play a role in determining the ground state here.  In particular,  we note the possibility of a CF exciton condensate\cite{Jain1996} resulting from interlayer-pairing of electron and hole \CFs ~\cite{Jain1996} 
(Fig. 3a,b), in analogy to the condensate of electron-hole pairs that is observed when electrons occupy half-filled LLs in these same double layers~\cite{Li.17a,Liu.17a}. We note that the even denominator  $\nu^{\dagger}$ states are overall stronger than the odd denominators, a behaviour that is strikingly similiar to the bare electron-hole bilayers where the exciton condensate at half filling of each layer is generally the most dominant. 


The odd integer $\nu^{\dagger}=1/3$ states (black circles in Fig. 2b) disperse vertically  with layer imbalance (solid black lines in Fig. 2b and highlighted by blue arrows in Fig.~2a, c and d) and show no correlation with behaviour expected for \CF states.  The state appearing at electron filling fraction $\nu=\frac{1}{6}$ ($\nu^{\dagger}=\frac{1}{3}$), together with the observed insensitivity to layer imbalance is consistent with the  exciton condensate described by the Halperin $\Phi_{333}$ wavefunction, which has been theoretically predicted to stabilize in double layer systems but not previously observed. 

By contrast, the FQHE states at $\nu_{i}=1/3$ and $2/3$ do not disperse at all with layer imbalance (green circles in Fig. 2b).  We note that no state is expected here in the  non-interacting \CF picture since this filling corresponds to the expected Fermi surface. Pairing instability of such Fermi surface has been the focus of theoretical discussion, potentially giving rise to non-Abelian incompressible states ~\cite{Scarola2001,Ardonne2002,Wen2010,Zaletel.15}. 

To distinguish among possible ground states we performed Coulomb drag measurements in similar DLG structure in which each layer is shaped into a Hall bar geometry (see method and SI for detailed information on device geometry).
In the Coulomb drag measurement, the Hall resistance of each layer exhibits a quantized plateau in the presence of an incompressible FQHE state, and the quantization value provides a topological invariant characteristic of the ground state order ~\cite{Renn1992,Jain.03}. 
For the  integer \CF filling states, marked by blue  vertical lines in Fig.~3c, we observe zero-valued longitudinal drag resistance simultaneous with (nearly) quantized Hall resistance on both drag and drive layers (Fig.~3c-e). In Fig. 3d,e we indicate the expected $R^{drag}_{xy}$ and $R^{drive}_{xy}$ plateau values according to the \CF construction (see SI) by horizontal blue lines.  We find good agreement with the measured values, providing further validation for the 2+1 flux CF interpretation of these states ~\cite{Renn1992}.

For the even denominator \CF states, (red vertical lines in Fig. 3c), the drag response demonstrates two types of behavior depending on filling fraction range.  At $\nu_{i} = -\frac{3}{7}$, where $\nu^{\dagger}=3/2$, the Hall resistance on both drive and drag layers approaches the expected plateau value for a composite exciton phase (red horizontal lines) with concomitant zero longitudinal drag resistance. Similar to the exciton condensate phase at $\nu_{i}=\frac{1}{2}$, such behavior in Coulomb drag measurement is determined by the balance between counterflow current of composite excitons in sample bulk and quasiparticle current along sample edge  (see SI for detailed calculation for composite exciton state) ~\cite{Li.17a,Liu.17a}. In the filling fraction range $-1< \nu_{i}<-1/2$,  boundary conditions require the current flow in the edge channel to be zero. Consequently both drive and drag layers are expected to exhibit insulating behaviour in the Coulomb drag geometry (see SI). Such insulating behavior is clearly demonstrated by the diverging resistance peak at $\nu_{i}=-\frac{4}{7}$, exceeding $100$ k$\Omega$ (Fig.~2c), with a similar resistance peak present at $\nu_{total}= -\frac{8}{13}$ and $-\frac{8}{11}$. Taken together, Coulomb drag measurement in matched density condition provides strong evidence for exciton pairing between \CFs ~\cite{Jain1996} when both graphene layers correspond to half-filled $\Lambda$-levels.

Finally, we briefly address the coloumb drag response for the other two types of non-integer \CF states. 
In Fig.~3c-e we observe a robust Hall resistance plateau at $\nu_i=\frac{2}{3}$, where a \CF Fermi surface is expected. The drive and drag layer Hall resistance  quantizes to h/e$^2$ and $\frac{1}{2}$h/e$^2$, respectively,  simultaneously with zero longitudinal resistance.  This rules out the possibility of this feature being a conventional Laughlin state of two decoupled graphene layers ~\cite{Eisenstein1992,Suen1994},  while providing direct evidence for ground state with interlayer correlation. A detailed study of this behaviour with varying interlayer separation and magnetic field may resolve the origin of this state. However, we note that at present there is no theory of Hall drag associated with the pairing states that have been proposed for this filling fraction ~\cite{Scarola2001,Ardonne2002,Wen2010,Zaletel.15}.  
The odd-denominator \CFs,  were not observable in our drag measurements (Fig. 3c-e), making it difficult to comment further beyond their observation in the Corbino measurement. We emphasize that the overall transport measurement resolution in the Hall bar geometry (Fig.~3) shows less resolution than our Corbino geometry (Fig.~1d), which is consistent with recent findings  in MLG devices ~\cite{Zeng2018}. Further investigation in improved hall bar geometries will likely be required to resolve nature of these states.

In summary, our results confirm the two component CF construction as a robust framework to model strongly interacting bilayer systems in the QHE regime. A sequence of the observed FQHE states are well described as effective IQHE states of CFs within the expanded two-component CF model.  Additional FQHE states, ocurring at fractional CF Landau filling, are consistent  with the formation of pair-interaction driven states, resulting from residual interlayer and intralayer interactions bewteen the CF quasi-particles. In particular half filled CF LLs in the double layer system appear to stabilize a ground state of indirect CF excitons, analagous to the electron-hole exciton formation observed at total integer electron filling in quantum Hall bilayers. 
Overall our observations suggest that the two-component CF construction for bilayers exhibits a self-similar correspondence to the bare electron bilayer behaviour but with residual interactions playing an important role.  We note that interlayer coupling between the CFs remains evidently strong, and tunable with effective layer separation providing a dynamic new way to study pairing between CFs. 
More generally we establish double layer graphene as a highly tunable  system for studying pairing interaction between quasi-particles, and pave the way for systematic examination of exotic phases with novel topological and statistical properties.

\section{method}
Charge carrier density in top and bottom layers, $n_{top}$ and $n_{bot}$, can be independently controlled by applying voltage bias to top and bottom graphite gate electrodes. Bulk conductance in parallel flow geometry is measured between the inner and outer edges of the corbino shaped sample while changing applied gate voltage on top and bottom graphite electrodes. Interlayer separation is $3$ nm for the device with corbino geometry, and $2.5$ nm in the device with Hall bar geometry.  

In a sample with aligned Hall bar geometry as shown in the inset of Fig.~3c, Coulomb drag measurement is performed by flowing current in one graphene layer (the drive layer) while the other layer (the drag layer) is grounded.
Longitudinal and Hall voltages are measured simultaneously on both graphene layers, which is made possible by shaping the double-layer structure into aligned Hall bar geometry and making electrical contact to each graphene layer independently. Current flows through one graphene layer (referred to as the drive layer) while the other layer (referred to as the drag layer) is kept at ground potential . Hall resistance on the drive layer is defined as $R_{xy}^{drive} = V_{xy}^{drive}/I_{drive}$, whereas longitudinal and Hall resistance on the drag layer are defined as $R_{xx}^{drag} = V_{xx}^{drag}/I_{drive}$ and $R_{xy}^{drag} = V_{xy}^{drag}/I_{drive}$, respectively. $V_{xy}^{drive}$, $V_{xx}^{drag}$ and $V_{xy}^{drag}$ are labeled in the inset of Fig.~3e. In the presence of ground state with interlayer correlation, Hall resistance in both drive and drag layers are expected to form well-defined plateau, with the plateau value directly related to the nature of interlayer correlation. At the same time zero longitudinal resistance is expected.

\begin{acknowledgments}
This work was supported by the National Science Foundation (DMR-1507788) and by the David and Lucille Packard Foundation. Data analysis was partially supported by US Department of Energy (DE-SC0016703). A portion of this work was performed at the National High Magnetic Field Laboratory, which is supported by National Science Foundation Cooperative Agreement No. DMR-1157490 and the State of Florida.

\end{acknowledgments}

\section*{Competing financial interests}
The authors declare no competing financial interests.

\newpage

\section{Supplementary Materials: \\ ``Novel composite fermion states with interlayer correlation in double-layer graphene''}

\section{Relationship between ($\nu_1$,$\nu_2$) and ($\nu_1^{\dagger}$,$\nu_2^{\dagger}$)}

For CF construction with $a=3$ and $b=1$, the relation between effective CF filling ($\nu_1^{\dagger}$,$\nu_2^{\dagger}$) and electron filling in each graphene layer ($\nu_1$,$\nu_2$) can be obtained by inverting Eq.2 in the main text,
\begin{align}
\nu_{i}=\frac{\nu_{i}^{\dagger}(1+\nu_{j}^{\dagger})}{1+2(\nu_{i}^{\dagger}+\nu_{j}^{\dagger})+3\nu_{i}^{\dagger}\nu_{j}^{\dagger}}\,.
\end{align}
\noindent
Further, constant $\nu_i^{\dagger}$ lies on a straight line in the $(\nu_1,\nu_2)$ phase space described by,
\begin{equation}
m\nu_{i}+\nu_{j}=1\,,
\end{equation}
where $m$ corresponds to the slope of the line in the $(\nu_1,\nu_2)$ phase space. Note that this straight line goes through $[0,1]$ and $[m^{-1},0]$, which is evident in Fig.~3 of the main text. More specifically, $m$ is defined as,
\begin{equation}
m^{-1} = \nu_{i}^{\dagger}/(2\nu_{i}^{\dagger}+1).
\end{equation} 
It has the same value as filling fraction of the Jain sequence of FQHE states in a single 2D confinement.

For $1 < \nu_{total} < 2$, CFs are constructed out of the hole type charge carriers and $\nu_i$ should be replaced by $1-\nu_i$.

\begin{figure*}
 \includegraphics[width=0.7\linewidth]{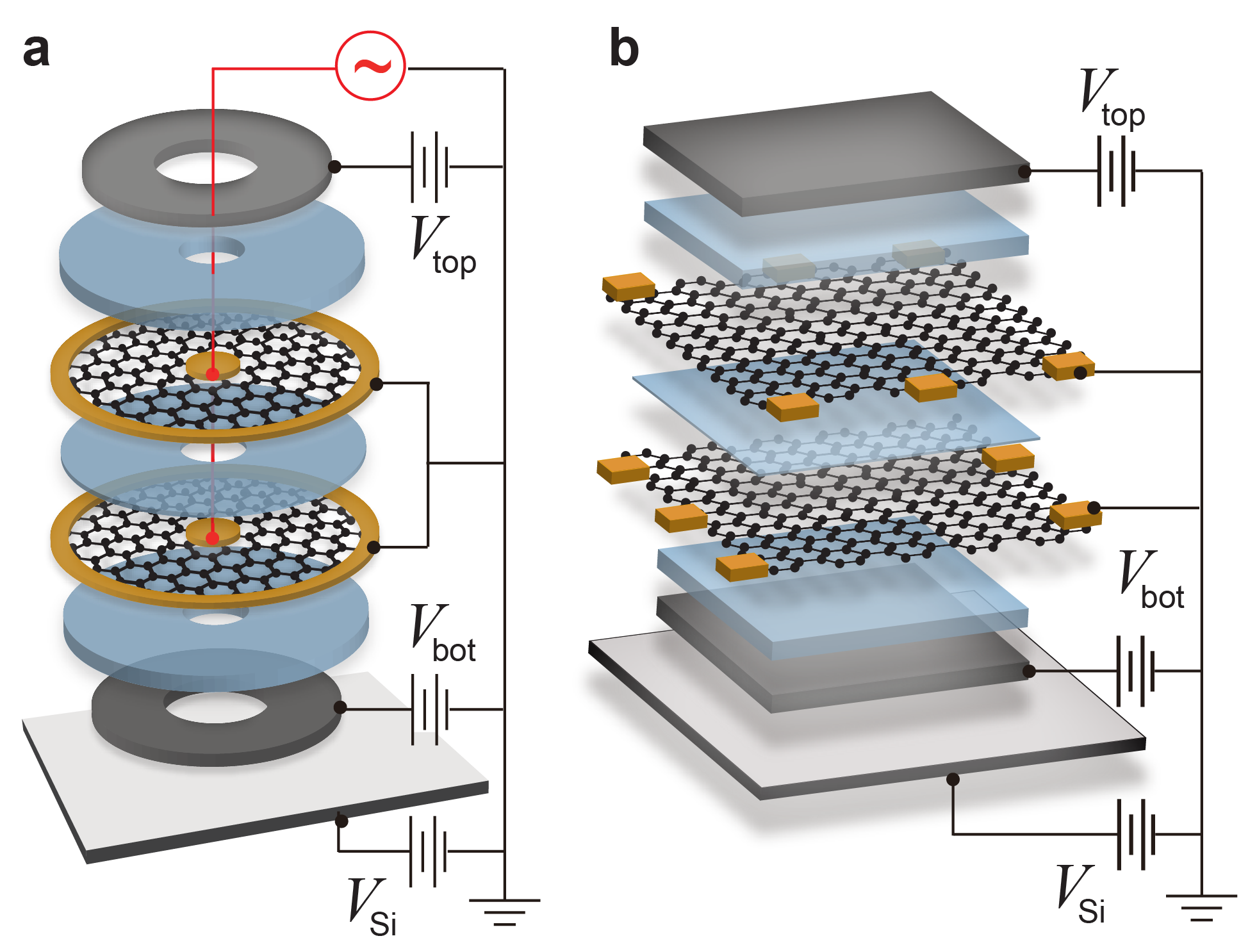}
 \caption{\label{fig4} {\bf{Device schematics}} (a) Schematic for device with Corbino geometry. In this work, the inner and outer edge of both graphene anulus are tied together. Bulk conductance is measured while flowing current through both graphene layers at the same time.     (b) Schematic for device with aligned Hall bar geometry. 1D edge contact each graphene layer separatedly. In the absence of interlayer tunneling, current only flows in the drive layer in Coulomb drag measurement, and net current flow in the drag layer is zero.}
\end{figure*}

\section{Hall resistance in Coulomb drag measurement: Integer $\nu^{\dagger}$}

In the CF model for a single 2D confinement, Hall resistance consists of two terms,
\begin{equation}
R_{xy}=R_{xy}^{CF}+R_{xy}^{CS}. \label{eqSI1}
\end{equation}
The CF term is related to the effective IQHE of CFs, $R_{xy}^{CF}= \frac{1}{\nu^{\ast}}\frac{h}{e^2}$, and the Chern-Simons (CS) term is associated with the transverse electric field generated by moving magnetic fluxes, $R_{xy}^{CS} = a \frac{h}{e^2}$, where $\nu^{\ast}$ is the effective filling for CFs, and $a$ is the number of flux attached to each electron in the CF construction ~\cite{Jain.03}. As a result, the Hall resistance plateau for fractional quantum Hall states are,
\begin{equation}
R_{xy}= \frac{a\nu^{\ast}+1}{\nu^{\ast}}\frac{h}{e^2}. \label{eqSI1b}
\end{equation}
Hall resistance for a double-layer structure in Coulomb drag measurement may be modelled in the CF model along the same line. In an aligned Hall bar device, the Hall voltage measured on both layers are related to vortex attachment parameters of $a$ and $b$, along with effective filling fraction $\nu^{\dagger}$ for two-component CFs ~\cite{Jain.03,Renn1992}. 
 We express the Hall resistance on both drive and drag layer in matrix form, where diagonal matrix element represents the drive layer response and off-diagonal element corresponds to drag layer response.
We will first discuss the situation along equal density line, where the effective filling fraction is the same across two graphene layers, $\nu_1^{\dagger}=\nu_2^{\dagger}\equiv\nu^{\dagger}$. Following Eq.~\ref{eqSI1}, we can similarly define the Hall resistance related to CF movement as the sum of CF and CS terms defined as the following:
\begin{equation}
R_{xy}^{CF}=
\frac{h}{e^2} 
\begin{bmatrix}
\frac{1}{\nu^{\dagger}} & 0 \\
0 & \frac{1}{\nu^{\dagger}}
\end{bmatrix}, \label{CF}
\end{equation}
and, 
\begin{equation}
R_{xy}^{CS}=
\frac{h}{e^2} 
\begin{bmatrix}
a & b \\
b & a
\end{bmatrix}. \label{CS}
\end{equation}
Combining Eq.~\ref{CF} and ~\ref{CS},

\begin{equation}
R_{xy}=\begin{bmatrix}
R_{xy}^{drive} & R_{xy}^{drag} \\
R_{xy}^{drag} & R_{xy}^{drive}
\end{bmatrix}
=
\frac{h}{e^2} 
\begin{bmatrix}
\frac{a\nu^{\dagger} + 1}{\nu^{\dagger}} & b \\
b & \frac{a\nu^{\dagger} + 1}{\nu^{\dagger}}
\end{bmatrix} 
=
\frac{h}{e^2} 
\begin{bmatrix}
m & b \\
b & m
\end{bmatrix} \label{eqSI4}
\end{equation} 
\noindent
In the last step the expression is simplified by defining $m^{-1}=\frac{\nu^{\dagger}}{a\nu^{\dagger} + 1}$, which corresponds to the filling fraction of CF states in a single layer, or the slope of constant effective filling line for $CF_2^1$. 
As an example, at filling fraction $\nu_{total}=\frac{4}{5}$, the ground state corresponds to $\nu^{\dagger}=-2$ of CFs with $a=2$ and $b=1$. According to Eq.~\ref{eqSI4}, Hall resistance on drive and drag layer should be $R_{xy}^{drive}=\frac{3}{2}\frac{h}{e^2}$ and $R_{xy}^{drag}=\frac{h}{e^2}$, respectively.

In filling fraction range $1 < \nu_{total} < 2$, electron-hole conjugate states coexist with an integer edge channel that also contributes to Hall conductance. The Hall resistance can be written as,
\begin{equation}
\Tilde{R}_{xy}=\Tilde{\sigma}_{xy}^{-1}=[e^2/h I - R_{xy}^{-1}]^{-1}\,
\label{eqSI6}
\end{equation} 
where $I$ is the identity matrix and $R_{xy}$ is the Hall resistance of the conjugate electron state given by Eq.~\ref{eqSI4}.
The diagonal and off-diagonal terms can be found as,
\begin{equation}
    R_{xy}^{drive}=\frac{m(m-1)-b^2}{(m-1)^2-b^2}\frac{h}{e^2}\,,
\end{equation}
and 
\begin{equation}
    R_{xy}^{drag}=\frac{-b}{(m-1)^2-b^2}\frac{h}{e^2}\,.
\end{equation}
At  $\nu_{total}=\frac{6}{5}$, ground state corresponds to the particle-hole conjugate of $\nu_{total}=\frac{4}{5}$, and the expected Hall resistance on both drive and drag layers are  $R_{xy}^{drive}=\frac{1}{3}\frac{h}{e^2}$ and $R_{xy}^{drag}=\frac{4}{3}\frac{h}{e^2}$.

Away from the equal density line, $\nu_1\neq\nu_2$, effective filling fraction differs in each graphene layer, $\nu_1^{\dagger}\neq\nu_2^{\dagger}$. In this case, combining generalized Eq.~\ref{CF}, Eq.~\ref{eqSI1} and Eq.~\ref{eqSI6} gives,
\begin{equation}
R_{xy}^{drag}=\frac{-b(m_1 m_2-b^2)}{[m_1(m_2-1)-b^2][m_2(m_1-1)-b^2]-b^2}\,. \label{neq2}
\end{equation}
where $m_i^{-1}=\frac{\nu_i^{\dagger}}{a\nu_i^{\dagger} + 1}$ with $i$ being the layer index. 
For $a=2$ and $b=1$, the numerator in Eq.~\ref{neq2} is always negative, while the denominator changes sign across the boundary corresponding to $\nu_1^{\dagger}+\nu_2^{\dagger}=-1$. Therefore a sign change in $R_{xy}^{drag}$ is expected across this boundary which is plotted as the black dashed line in Fig.4d of the main text and confirmed by measured Hall drag signal shown in Fig.~4e.

\section{Coulomb drag measurements at Half-integer $\nu^{\dagger}$}

At half-integer $\nu^{\dagger}$, the incompressible states could result from paring of CFs in the 2 layers. 
Because of the CF construction still applies, the Hall resistance could be written as the sum of CF and CS terms as discussed above, with $R_{CS}$ given by Eq.~\ref{CS}.
However, $R_{CF}$ is modified as,
\begin{equation}
R_{xy}^{CF}=
\frac{h}{e^2}\frac{1}{2\nu^{\dagger}}
\begin{bmatrix}
1 & 1 \\
1 & 1
\end{bmatrix}. \label{FEC}
\end{equation}
The equal diagonal and off-diagonal elements is a result of CF paring, and in analogy to the Coulomb drag in the presence of excitons formed by electrons and holes when $\nu_1 + \nu_2 =$ integer.
For example, filling fraction $\nu_{total}=\frac{6}{7}$ corresponds to $\nu^{\dagger}=-3/2$ of CFs with $a=2$ and $b=1$. Combining Eq.~\ref{FEC} with Eq.~\ref{eqSI1} and Eq.~\ref{CS}, we get $R_{xy}^{drive}=\frac{5}{3}\frac{h}{e^2}$ and $R_{xy}^{drag}=\frac{2}{3}\frac{h}{e^2}$.

For $1 < \nu_{total} < 2$, an additional integer edge channel contributes to the conductance.
Following Eq.~\ref{eqSI6}, it can be found that $\Tilde{\sigma}_{xy}^{drive} = \Tilde{\sigma}_{xy}^{drag}$, which further gives $I^{drive}=I^{drag}$.
Since $I^{drag} = 0$ in the experimental setup, the system is insulating and $R_{xx}$ and $R_{xy}$ in both drive and drag layers should diverge. 
The abnormally high $R_{xx}^{drag}$ at $\nu_{tot}=8/7$ as shown in Fig.4 in the main text is thus consistent with the model of CF pairing.

\section{Effective half filling state away from the equal density line}

The effective half-filling state could result from exciton pairing between composite electron and composite holes when both layers are tuned to half-filled $\Lambda$-levels. 
However, a potential composite exciton ground state is not expected to be stable away from the equal density line. This is illustrated in Fig.~\ref{Dnu}.

Electron filling in each graphene layer is determined by the ratio between the number of charge carrier and magnetic flux, $\nu_i=n_i\phi_0/B$, where $i$ is the layer index and $\phi_0$ the quantum of magnetic field, or a flux quantum. Red (blue) arrows represent magnetic flux quanta penetrating top (bottom) graphene layer, some of which are adsorbed by CFs to model electron correlation both within the same layer and between two graphene layers. Total number of magnetic flux remains the same for top and bottom graphene layers after the formation of CF, but CFs only experience an effective magnetic field determined by the number of unattached flux, $B_i^{\dagger}/\phi_0$, which is reduced from $B/\phi_0$. Notice that in the presence of density imbalance where $\nu_1 \neq \nu_2$, the effective magnetic field is different for two graphene layers. For CFs with $2$ intralayer and $1$ interlayer flux attachment, the residual magnetic field is $B_{i}^{\dagger} = B - (2 n_i+ n_j)\phi_0$, and the effective filling
fraction of $\Lambda$-level is defined as the ratio between the number of CFs and residual magnetic flux, $\nu_i^{\dagger}=n_i\phi_0/B_i^{\dagger}$.

In Fig.~\ref{figDnu}b, effective CF filling is half integer in both graphene layers with $\nu_1^{\dagger}=1/2$ and $\nu_2^{\dagger}=3/2$. However, since the effective magnetic field is different, $B_1^{\dagger}= 10\phi_0 $ and $B_2^{\dagger}= 6\phi_0 $, the $\Lambda$-level degeneracy is different, which leads to different numbers of composite electron and composite holes on different graphene layers, making an exciton ground state unstable.

It is similarly shown in Fig.~\ref{figDnu}c for effective CF fillings of $\nu_1^{\dagger}=1/3$ and $\nu_2^{\dagger}=2/3$ in two graphene layers. It is evident that exciton ground state is not stable as the number of composite electrons and composite holes are different on two graphene layers. 

The comparison between Fig.~\ref{figDnu}b and c also reveals another subtlty of the effective half filling states in a double-layer structure. An exciton ground state of electron-hole pair is robust against density imbalance and tracks constant total filling fraction of $\nu_{total}=1$, where the number of electrons and holes are the same across two graphene layers. The trajectory of the effective half filling states in Fig.~3b of main text do not track constant total filling of CFs, as shown in Fig.~\ref{figDnu}c, instead it is follows constant CF filling of a single graphene layer. In another word, the robustness of the effective half filling state in one graphene layer is independent of the opposite layer . This suggests that the effective half filling state away from the equal density line is a result of correlation between CFs within a single layer, \emph{i.e.} a Pfaffian or $\phi_{331}$ state of CFs. It is equally important to note that this does not rule out the possibility of an exciton ground state at effective half filling along the equal density line.

\begin{figure*}
 \includegraphics[width=1\linewidth]{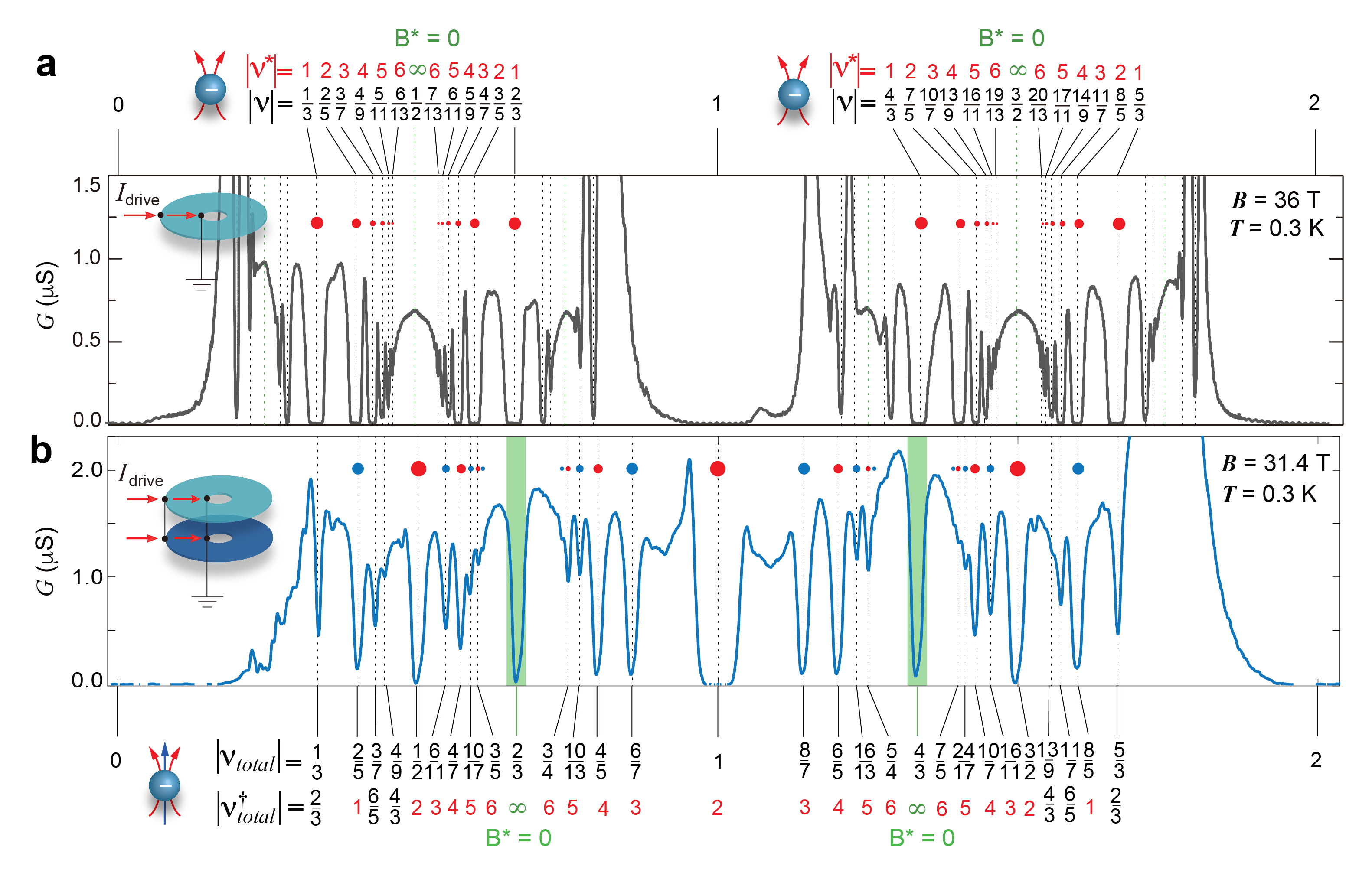}
 \caption{\label{fig1}{\bf{Comparing bulk conductance of graphene monolayer and double-layer.}} (a) Linecut of bulk conductance $G$ versus $\nu$ measured at $B = 36$ T in monolayer graphene device with Corbino geometry ~\cite{Zeng2018}. Both electron filling fraction $\nu$ and effective fillings $\nu^{*} $ of the 2-flux CFs ($CF_2^0$ ) are shown on the top axis for each conductance minimum. The size of the red circle is proportional to the strength of energy gap associated with the conductance minimum.  (b) Linecut of $G$ versus $\nu_{total}$ along equal density line measured at $B = 31.4$ T. Total electron filling across two layers are varied between $0< \nu_{total} < 1$. The size of the red and blue circles is proportional to the energy gap associated with ground states at effective integer and half integer CF filling fractions. Both panel (a) and (b) are plotted in total filling fraction range $0 < \nu_{total} < 2$. The different ground state hierarchy in panel (b) compared to panel (a) demonstrates the effect of layer degree of freedom in graphene double-layer and is consistent with nagligible interlayer tunneling.  }
\end{figure*}

\begin{figure*}
 \includegraphics[width=1\linewidth]{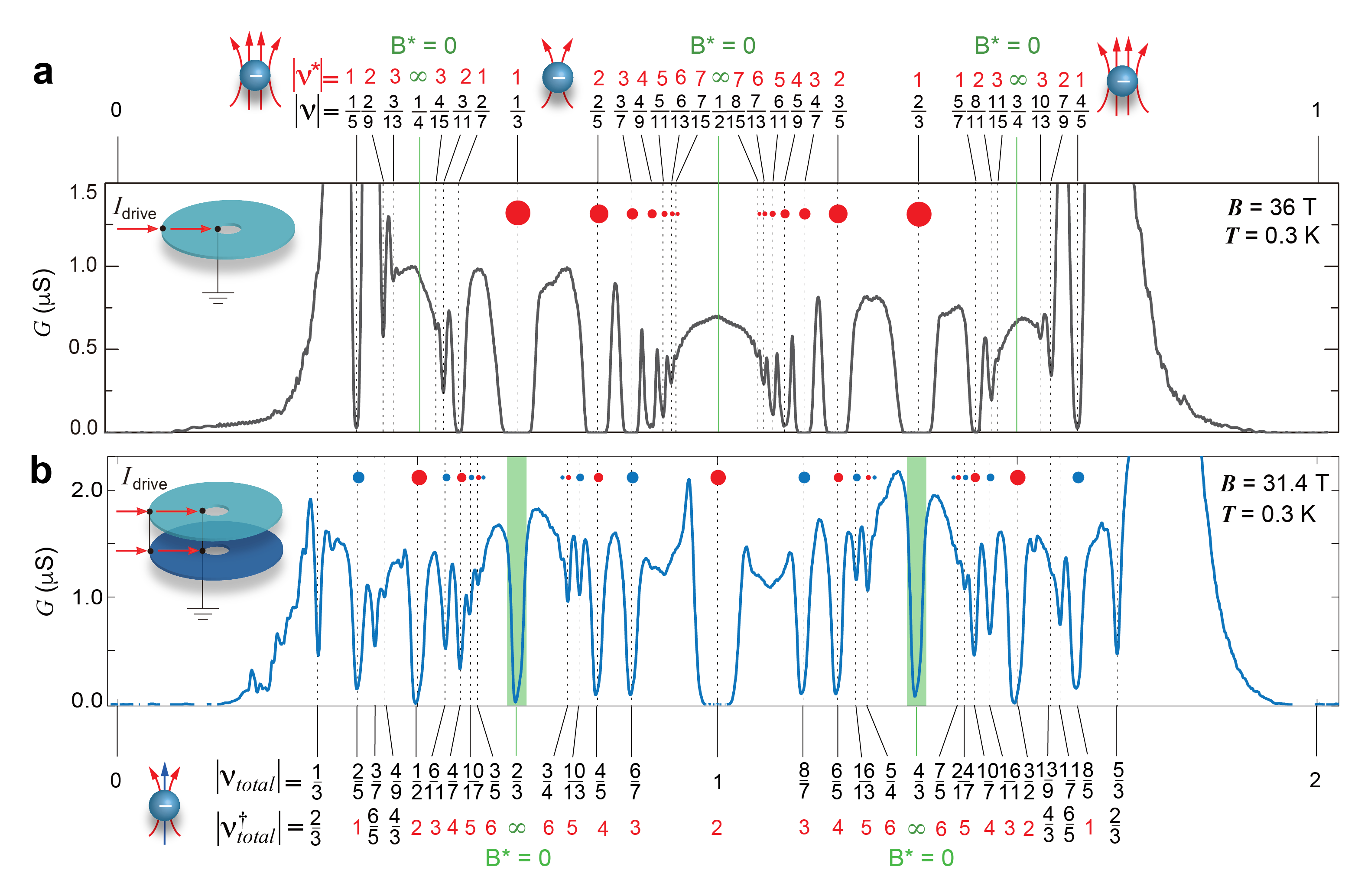}
 \caption{\label{fig2} {\bf{Comparing bulk conductance of graphene monolayer and double-layer.}} (a) Linecut of bulk conductance $G$ versus $\nu$ measured at $B = 36$ T in monolayer graphene in Corbino geometry in filling fraction range $0 < \nu < 1$ ~\cite{Zeng2018}. Both electron filling fraction $\nu$ and effective fillings $\nu^{*} $ of the 2-flux CFs ($CF_2^0$ ) are shown on the top axis for each conductance minimum. The size of the red circle is proportional to the strength of energy gap associated with the conductance minimum.  (b) Linecut of $G$ versus $\nu_{total}$ along equal density line measured at $B = 31.4$ T. Total electron filling across two layers are varied between $0< \nu_{total} < 1$. The size of the red and blue circles is proportional to the energy gap associated with ground states at effective integer and half integer CF filling fractions. Both panel (a) and (b) are plotted in filling fraction range $0 < \nu < 1$ of a single graphene layer, \emph{i.e.} total filling up to $1$ in graphene monolayer and $2$ in graphene double-layer. Different ground state hierarchy in panel (b) compared to panel (a) demonstrates that the single layer CF model failed to account for electron correlation in the double-layer and is evidence for interlayer correlation.  }
\end{figure*}

\begin{figure*}
 \includegraphics[width=1\linewidth]{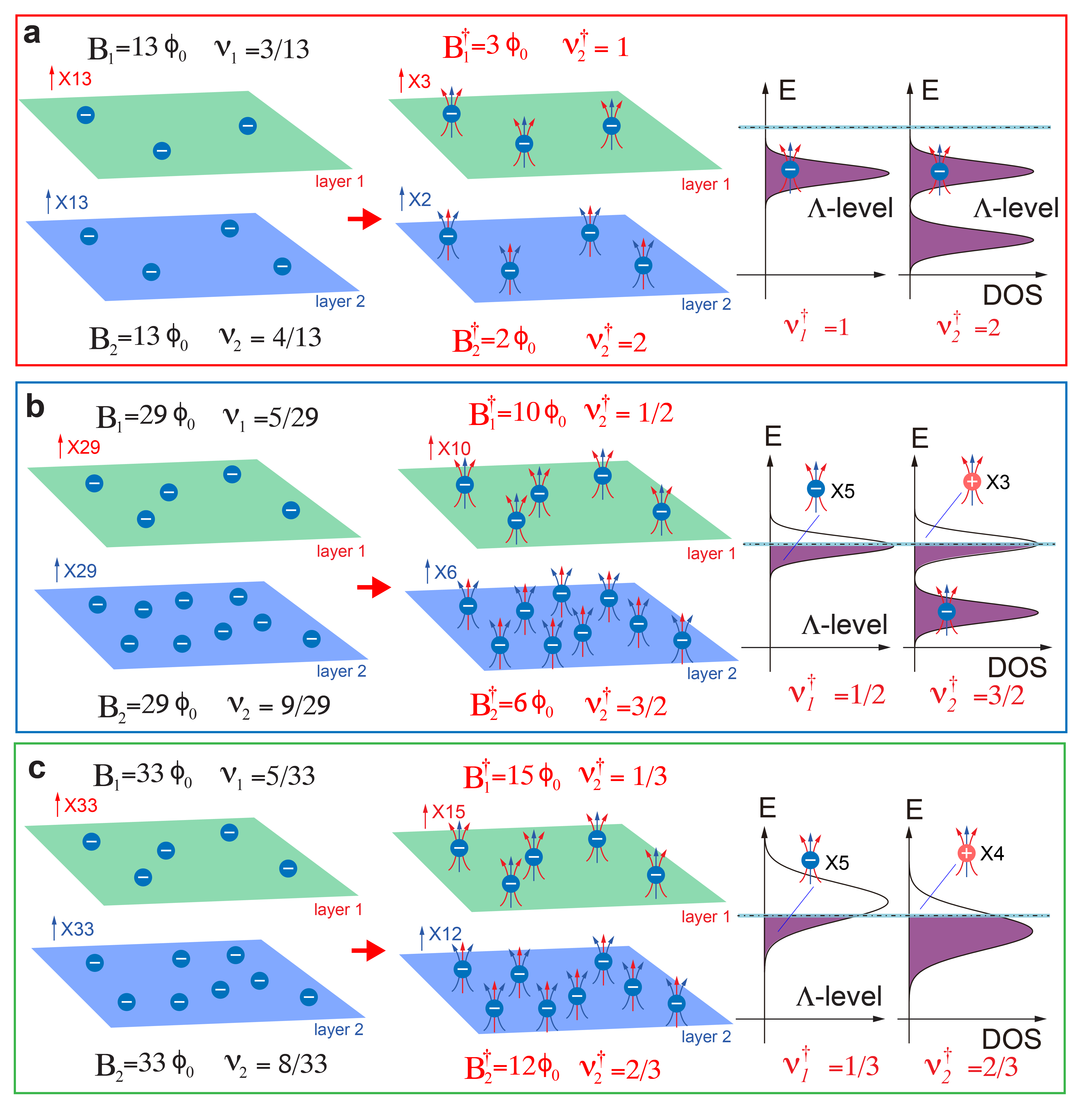}
 \caption{\label{figDnu} {\bf{Effective CF filling in the presence of density imbalance.}}  (a) Left panel, electron filling of $\nu_1=3/13$ and $\nu_2=4/13$. Middle panel, after flux attachment, effective CF filling becomes $\nu_1^{\dagger}=1$ and $\nu_2^{\dagger}=2$. Right panel, schematic of density of states (DOS) for CFs as a function of energy. $\Lambda$-levels of CFs are illustrated as a Gaussian distribution in DOS. At this filling fraction, fermi energy lies above the lowest $\Lambda$-level for layer $1$ and above the second lowest $\Lambda$-level for layer $2$. The ground states in each graphene layer is incompressible, \emph{i.e.}, sample bulk is insulating with a FQHE edge channel on sample edge. (b) Left panel, electron filling of $\nu_1=5/29$ and $\nu_2=9/29$. Middle panel, after flux attachment, effective CF filling becomes $\nu_1^{\dagger}=1/2$ and $\nu_2^{\dagger}=3/2$. Right panel, schematic of density of states (DOS) for CFs as a function of energy where $\Lambda$-level is half filled in both graphene layers. Notice that the residual magnetic field along with $\Lambda$-level degeneracy are different in two graphene layers, this leads to different numbers of composite electron and composite holes on different graphene layers, making an exciton ground state unstable. (c) Similar to (b), while the sum of effective filling across two layers is $1$, the number of composite electrons and composite holes is different on two graphene layers, making an exciton ground state unstable.  }
  \label{Dnu}
\end{figure*}

\begin{figure*}
 \includegraphics[width=0.95\linewidth]{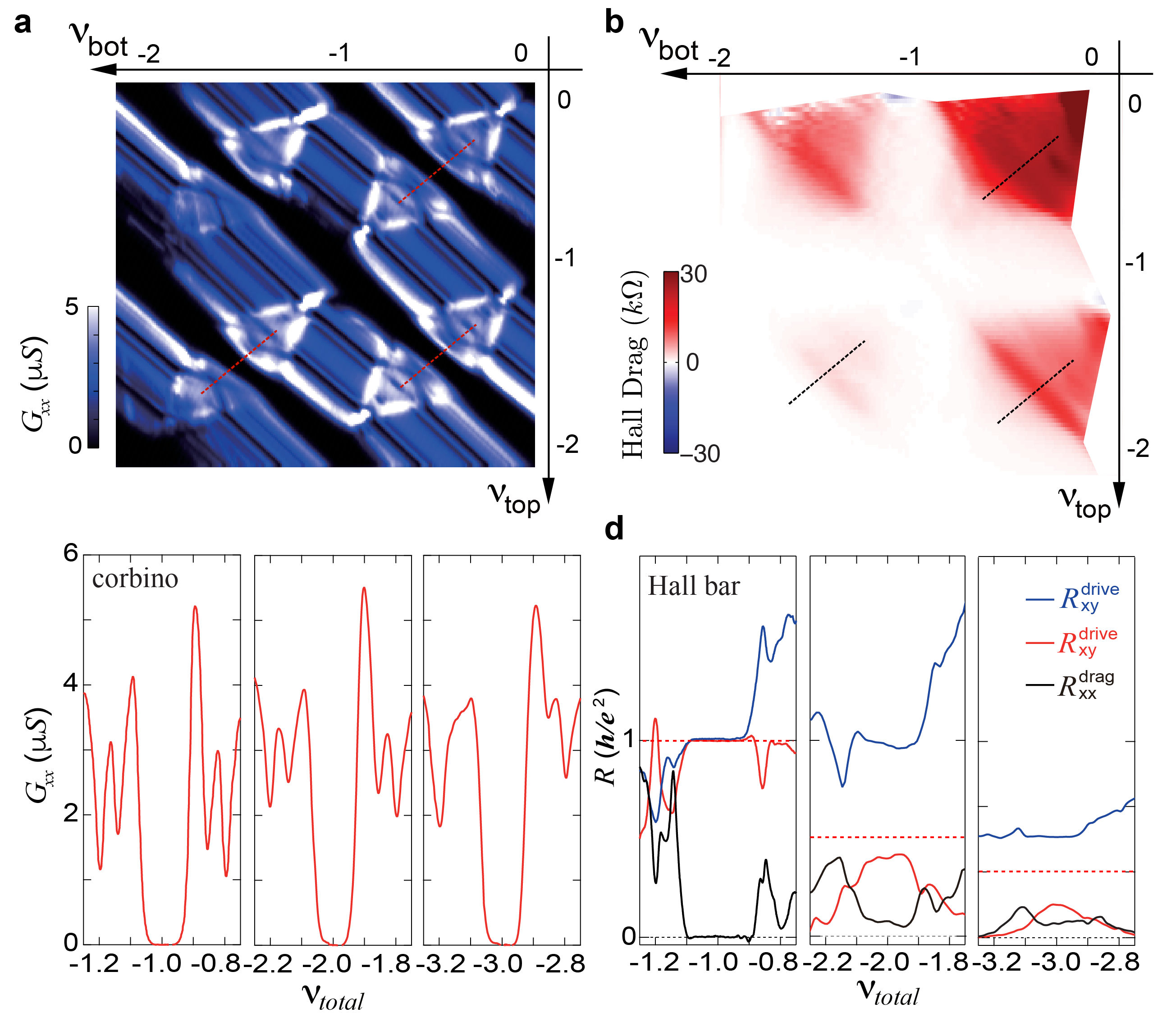}
 \caption{\label{figSI4} {\bf{Comparison between Corbino and aligned Hall bar geometry.}} Bulk conductance (a) and Hall drag response (b) as a function of filling fraction in top and bottom graphene layers, measured at $B = 15$ T and $T = 0.3$ K. (c) Linecuts of bulk conductance along red dashed lines in (a), as a function of $\nu_{total}$. Bulk conductance reaches zero at integer values of $\nu_{total}$, indicating a fully developed exciton condensate in sample bulk. (d) Linecuts of magneto-resistance in Coulomb drag geometry measured in device with aligned Hall bar geometry. Linecuts were taken along black dashed lines in (b), and ploted as a function of $\nu_{total}$. Around $\nu_{total}=-1$, Hall resistance of both drive and drag layers quantize to the same value while zero longitudinal resistance is observed, consistent with an exciton condensate ground state. For higher total filling, Coulomb drag measurements are indicative of an exciton ground state that is not fully developed. }
\end{figure*}

\end{document}